\documentclass[acmsmall,nonacm,11pt]{acmart}
\makeatletter\let\@authorsaddresses\@empty\makeatother
\usepackage[utf8]{inputenc}
\usepackage[T2A,T1]{fontenc}
\usepackage{natbib} % for \citep and \citet
\usepackage{csquotes}
\usepackage{adjustbox}
\usepackage{proof} % for \infer inference trees with nested levels
\usepackage{mdframed} % for drawing frames around figures
\usepackage[novert,noframes,nonumbers]{ffcode} % for fixed-fonts
\usepackage{CJKutf8} % for chinese font
\usepackage{multicol} % for multicolumn printing
\usepackage{paralist} % for inlined lists
\usepackage{anyfontsize} % To get rid of font not found warnings
\usepackage{eolang} % for EO sources and formulas
\usepackage{tabularx} % for special tables
\usepackage{naive-ebnf} % for EBNF
\usepackage{to-be-determined} % for \tbd command
\usepackage{href-ul} % for nicely underscored links
\usepackage[shortlabels]{enumitem} % for changing labes in "enumerate"
\usepackage[capitalize,nameinlink]{cleveref} % must stay last!
  
  \crefname{lemma}{Lemma}{Lemmas}
  \Crefname{lemma}{Lemma}{Lemmas}

\renewcommand\emph[1]{\textit{#1}} % to prevent a weird bug coming from some of the packages above
\newcommand\capt[1]{\caption{#1}\Description{#1}}

\newcommand\nospell[1]{#1}

\newcommand\phiTransNormalize{\mathrel{\overset{\raisebox{-1.0ex}{\(\scriptstyle *\)}}{\phiNormalize}}}

\def\nf{n}
\def\dead{\bot}

\definecolor{rgbTerminals}{HTML}{ff8c00}
\renewcommand\EbnfTerminal[1]{\textcolor{rgbTerminals}{#1}}
\renewcommand\phiTerminal[1]{\EbnfTerminal{#1}}

\newcommand\phiText[1]{\;\text{\textcolor{gray}{#1}}\;}

\makeatletter
\newcommand\phiOver[1]{%
  {\color{gray}\overbracket[0.4pt][1pt]{{\color{black}#1}}}}
\makeatother

\makeatletter
\def\eolangversion@parse#1 #2 #3\relax{#2}
\newcommand\eolangversion{\expandafter\expandafter\expandafter\eolangversion@parse\csname ver@eolang.sty\endcsname\relax}
\makeatother

\newread\phinoversionfile
\begingroup
\openin\phinoversionfile=phino-version.txt
\readline\phinoversionfile to \phinoversionline
\closein\phinoversionfile
\xdef\phinoversion{\phinoversionline}
\endgroup
\newcommand\phinotex{_tex/\phinoversion}
\iexec[maybe]{mkdir -p \phinotex}

\newcounter{phimeet}
\newcommand\phinoKey[1]{{\def\mathrel##1{##1}\def\phiTerminal##1{##1}\xdef\phinoKeyValue{#1}}}
\makeatletter
\newcommand\phinoMeet[2]{%
  \ifmeasuring@\else\refstepcounter{phimeet}\phinoKey{#1}\cref@label{phimeet:\phinoKeyValue}\fi%
  \def\param{#1}%
  {\color{gray}\overbracket[0.4pt][1pt]{{\color{black}#2}}^{{\color{gray}\Pi_{\thephimeet}}}}}
\makeatother
\crefformat{phimeet}{{\color{gray}\Pi_{#2#1#3}}}
\newcommand\phinoAgain[1]{\phinoKey{#1}\cref{phimeet:\phinoKeyValue}}

\newcounter{rule}
\makeatletter
\newcommand\rulemnemo[1]{\text{\sffamily{}R\(_\text{\scshape\sffamily{}#1}\)}}
\newcommand\newrule[2][]{%
  \def\param{#1}%
  \ifx\param\empty%
    \def\sub{#2}%
  \else%
    \def\sub{\(#1\)}%
  \fi%
  \refstepcounter{rule}%
  \protected@edef\@currentlabelname{\rulemnemo{\sub}}%
  \label@noarg{r:#2}%
  \rulemnemo{\sub}%
  }
\makeatother

\newcommand\phinoNormalizationRule[6][]{%
  \newrule[#1]{#2}:
  &
  \phiq{\phinoNormalize{#3}{#4}}
  \ifblank{#5}{}{ \phiq{\phiText{if} #5} }%
  \\
  \ifblank{#6}{}{ & \phiq{\phiText{if} #6} \\ }%
  }
\newcommand\phinoEvaluate[5]{%
  \langle #1 {,}\; #2 {,}\; #3 \rangle \phiEvaluate{} \langle #4 {,}\; #5 \rangle}
\newcommand\phinoNormalize[2]{%
  #1 \phiNormalize{} #2}
\newcommand\phinoDataize[5]{%
  \langle #1 {,}\; #2 {,}\; #3 \rangle \phiDataize{} \langle #4 {,}\; #5 \rangle}
\newcommand\phinoMorph[5]{%
  \langle #1 {,}\; #2 {,}\; #3 \rangle \phiMorph{} \langle #4 {,}\; #5 \rangle}
\newcommand\phinoContextualize[3]{%
  \langle #1 {,}\; #2 \rangle \phiContextualize{} #3}

\newcommand\phinoNotFormation[1]{#1 \notin \mathcal{B}}
\newcommand\phinoNotAbsolute[1]{#1 \notin \mathcal{K}}

\newcommand\phinoNamed[2]{#1 = #2}

\makeatletter
\let\phino@premises\@empty
\newcommand\phinoName[1]{\gdef\phino@name{#1}}
\newcommand\phinoLabel[1]{\gdef\phino@label{#1}}
\newcommand\phinoConclusion[1]{\gdef\phino@conclusion{\phiq{#1}}}
\newcommand\phinoCondition[1]{\gdef\phino@condition{\phiq{#1}}}
\newcommand\phinoPremise[1]{%
  \ifx\phino@premises\@empty
    \gdef\phino@premises{\phiq{#1}}%
  \else
    \g@addto@macro\phino@premises{\qquad \phiq{#1}}%
  \fi}
\newenvironment{phinoContextualizationInference}
  {\begin{phinoInference}}
  {\end{phinoInference}}
\newenvironment{phinoMorphingInference}
  {\begin{phinoInference}}
  {\end{phinoInference}}
\newenvironment{phinoDataizationInference}
  {\begin{phinoInference}}
  {\end{phinoInference}}
\newenvironment{phinoInference}{%
  \global\let\phino@name\@empty
  \global\let\phino@label\@empty
  \global\let\phino@premises\@empty
  \global\let\phino@conclusion\@empty
  \global\let\phino@condition\@empty
}{%
  \adjustbox{max width=\linewidth,margin=0em .3em}{%
    \ifx\phino@name\@empty\else%
      \ifx\phino@label\@empty%
        \newrule{\phino@name}\;%
      \else%
        \newrule[\phino@label]{\phino@name}\;%
      \fi%
    \fi%
    \ensuremath{\dfrac{\text{\phino@premises}}{\text{\phino@conclusion}}}%
    \ifx\phino@condition\@empty\else\phiText{if}\phino@condition\fi%
  }}%
\makeatother

\acmBooktitle{untitled}
\title{\texorpdfstring{\(\upvarphi\)}{phi}-Calculus: Object-Oriented Formalism}
\subtitle{%
  Ver:
  \texorpdfstring{
    \href{https://github.com/objectionary/calculus-paper/releases/tag/0.11.0}
      {\ff{0.11.0}}
  }{0.11.0}
}
\author{Yegor Bugayenko}
  \orcid{0000-0001-6370-0678}
  \email{yegor256@gmail.com}
  \affiliation{\institution{Huawei}\city{Moscow}\country{Russia}}
\author{Maxim Trunnikov}
  \orcid{0000-0001-8545-4797}
  \email{mtrunnikov@gmail.com}
  \affiliation{\institution{Huawei}\city{Moscow}\country{Russia}}


\begin{document}

% SPDX-FileCopyrightText: Copyright (c) 2016-2026 Objectionary.com
% SPDX-License-Identifier: MIT

\begin{abstract}
Object-oriented programming (OOP) is one of the most popular paradigms used for building software systems%
  \footnote{%
  \LaTeX{} sources of this paper are maintained in the
  \href{https://github.com/objectionary/calculus-paper}{objectionary/calculus-paper} public GitHub repository,
  the rendered version is \href{https://github.com/objectionary/calculus-paper/releases/tag/0.11.0}{\ff{0.11.0}}.}.
However, despite its industrial and academic popularity,
  OOP is still missing a formal apparatus similar to \(\lambda\)-calculus,
  which functional programming is based on.
A number of attempts were made to formalize OOP, but none of them managed
  to cover all the features available in modern OO programming languages, such as C++ or Java.
We have made yet another attempt and created \phic{}, intended as a minimal
  foundation on which such features may be built rather than a calculus that
  already covers them.
This paper does not demonstrate the practical use or effect of \phic{} but merely explains it.
\end{abstract}

\maketitle

% SPDX-FileCopyrightText: Copyright (c) 2016-2026 Objectionary.com
% SPDX-License-Identifier: MIT

\section{Introduction}

Object-oriented programming (OOP) has become
  the dominant paradigm for software development,
  with languages such as Java, C++, Python, and C\#
  being among the most widely used in industry%
  ~\citep{madsen1988object,nierstrasz1989survey,korson1990understanding,booch1994object,meyer1997object,west2004object}.
Despite this prevalence, the theoretical foundations of OOP remain
  less developed than those of functional programming%
  ~\citep{rentsch1982object,madsen1988object},
  which has long benefited from the
  \(\lambda\)-calculus~\citep{church1936some,barendregt2012} as a formal basis.
The absence of a comparable formalism for OOP hinders rigorous reasoning
  about object-oriented programs,
  complicates the development of verified compilers and static analyzers,
  and limits the potential for principled language design%
  ~\citep{eden2001principles}.

Several attempts have been made to formalize object-oriented (OO) concepts.
\citet{abadi1995imperative} proposed an imperative object calculus,
  while \citet{igarashi2001featherweight} introduced Featherweight
  Java as a minimal core calculus.
However, these formalisms either focus on specific language features
  or omit constructs commonly found in modern OO languages.
As \citet{nierstrasz1991towards} observed, the development
  of concurrent object-based programming languages
  has suffered from the lack of any generally accepted formal foundations.

This paper presents \phic{}%
  \footnote{Typeset with the \ff{eolang} \LaTeX{} package, version \ff{\eolangversion}.},
  a formalism designed to serve
  as a foundation for object-oriented programming.
The calculus is built around a primitive---the object formation---which
  encapsulates formations, data, and functions.
Objects interact through application, which attaches expressions to attributes,
  and dispatch, which retrieves attached expressions.

The contributions of this paper are as follows:
\begin{itemize}
  \item A formal syntax for \phic{} defined
  by a context-free grammar (\cref{sec:syntax}).
  \item Rigorous definitions of the core constructs (\cref{sec:foundations}).
  \item A family of semantic operators (\cref{sec:operators}).
  \item A reference interpreter that runs
  \(\upvarphi\)-expressions\footnote{\url{https://github.com/objectionary/phino}}.
\end{itemize}

The remainder of this paper is organized as follows.
\Cref{sec:overview} provides an informal overview of the calculus.
\Cref{sec:syntax} presents the syntax as a context-free grammar.
\Cref{sec:foundations} defines the foundational concepts.
\Cref{sec:operators} introduces the semantic operators.
\Cref{sec:related} discusses related work.
Finally, \cref{sec:future} suggests directions of future studies.

% SPDX-FileCopyrightText: Copyright (c) 2016-2026 Objectionary.com
% SPDX-License-Identifier: MIT

\section{Informal Overview}\label{sec:overview}

An object \emph{formation} is a collection of \emph{attributes},
  which are uniquely named pairs, for example:
\begin{phiquation}
\label{eq:price-color}
[[ price -> ?, color -> [[ D> FF-C0-CB ]] ]] {.}
\end{phiquation}
This formation has two attributes \ff{price} and \ff{color}.
The formation is an \emph{abstract formation} because the \ff{price} attribute
  is \emph{void}, i.e. there is nothing \emph{attached} to it.
The \ff{color} attribute is attached to another formation
  with one $D$-\emph{asset},
  which is attached to \emph{data} (three bytes).

A \emph{full application} of a pair---an attribute and
  an \emph{expression}---to an abstract formation
  \emph{results} in a new \emph{closed} formation, for example:
\begin{phiquation}
\label{eq:simple-application}
[[ |x| -> ? ]]( |x| -> b ) \phiNormalize [[ |x| -> b ]] {.}
\end{phiquation}
A \emph{partial} application results in a new abstract formation, for example:
\begin{phiquation*}
[[ |x| -> ?, |y| -> ? ]]( |x| -> b ) \phiNormalize [[ |x| -> b, |y| -> ? ]] {.}
\end{phiquation*}

A formation may be \emph{dispatched} from another formation
  with the help of \emph{dot notation}
  where the right side \emph{accesses} the left side, for example:
\begin{phiquation}
\label{eq:dot-notation}
[[ |x| -> \phiOver{$.|p|(|t| -> b)}.|y|, |p| -> [[ |t| -> ?, |y| -> $.|t| ]] ]].|x| \phiNormalize b {.}
\end{phiquation}
The leftmost symbol \phiTerminal{\(\upxi\)} denotes the formation itself;
  the following $.|p|$ part retrieves the formation attached to the attribute \(\ff{p}\);
  then, the application $(|t|->b)$ makes a copy of the formation;
  finally, the $.|y|$ part retrieves formation \(b\).

Attributes are \emph{immutable}, i.e. an application to a formation,
  where the attribute is already attached,
  results in a \emph{terminator} denoted as $T$ (error), for example:
\begin{phiquation*}
[[ |x| -> b_1 ]]( |x| -> b_2 ) \phiNormalize T {.}
\end{phiquation*}

An expression may be textually reduced, for example:
\begin{phiquation*}
[[ |x| -> ?, |y| -> ? ]]( |x| -> b_1 )( |y| -> b_2 ) \phiNormalize
  \phiNormalize [[ |x| -> b_1, |y| -> ? ]] ( |y| -> b_2 ) \phiNormalize
  \phiNormalize [[ |x| -> b_1, |y| -> b_2 ]] {.}
\end{phiquation*}
The formation on the left is reduced to the formation on the right in two reduction steps.
Some expressions may be in \emph{normal form}, which means no further applicable reductions.

A formation is called a \emph{decorator} if it has the $@$-attribute
  with an expression attached to it,
  known as a \emph{decoratee}.
An attribute dispatched from a decorator reduces
  to the same attribute dispatched from the decoratee,
  if the attribute is not present in the decorator, for example:
\begin{phiquation*}
[[ @ -> [[ |x| -> b ]] ]].|x| \phiNormalize b {.}
\end{phiquation*}

A formation may have data attached only to its $D$-asset, for example:
\begin{phiquation*}
[[ |x| -> [[ D> |00-2A| ]] ]] {.}
\end{phiquation*}

A formation may have a \emph{function} attached to its $L$-asset.
Such a formation is referred to as an \emph{atom}.
An atom may be \emph{morphed} to another expression by evaluating its function,
  for example:
\begin{phiquation}
\label{eq:Sqrt}
\phinoMorph{[[ L> |Sqrt|, |x| -> 256 ]]}{e}{s}{[[ D> |10-| ]]}{s} {.}
\end{phiquation}
An expression may be \emph{dataized} by morphing its atoms first and then
  extracting the data attached to the $D$-asset of the formation,
  for example:
\begin{phiquation*}
\phinoDataize{[[ L> |Sqrt|, |x| -> 256 ]]}{e}{s}{|16|}{s} {.}
\end{phiquation*}

% SPDX-FileCopyrightText: Copyright (c) 2016-2026 Objectionary.com
% SPDX-License-Identifier: MIT

\section{Syntax}\label{sec:syntax}

The syntax is defined by BNF in \cref{fig:ebnf},
  where the starting symbol is \EbnfNonTerminal{Expression}.

\begin{figure*}
\begin{mdframed}
\raggedright
\begin{ebnf}[8em]
<Expression> := <Formation> | <Application> | <Dispatch> | <Locator> | "\(\dead\)" \\
<Formation> := "\(\llbracket\)" <Binding> "\(\rrbracket\)" \\
<Application> := <Expression> "\(\lparen\)" <\(\tau\)-Pair> "\(\rparen\)" \\
<Dispatch> := <Expression> "." <Attribute> \\
<Locator> := "\(\Phi\)" | "\(\upxi\)" \\
<Binding> := <Pair> <Bindings> | \(\epsilon\) \\
<Bindings> := "," <Pair> <Bindings> | \(\epsilon\) \\
<Pair> := <\(\varnothing\)-Pair> | <\(\tau\)-Pair> | <\(\Delta\)-Pair> | <\(\uplambda\)-Pair> \\
<\(\varnothing\)-Pair> := <Attribute> "\(\mapsto\)" "\(\varnothing\)" \\
<\(\tau\)-Pair> := <Attribute> "\(\mapsto\)" <Expression> \\
<\(\Delta\)-Pair> := "\(\Delta\)" "\(\phiDotted\)" <Data> \\
<\(\uplambda\)-Pair> := "\(\uplambda\)" "\(\phiDotted\)" <Function> \\
\end{ebnf}
\end{mdframed}
\capt{Syntax as a context-free grammar, in BNF.}
\label{fig:ebnf}
\end{figure*}

Besides the literals mentioned in the grammar in orange,
  the alphabet includes three non-terminals that rewrite to terminals as follows:
\begin{itemize}
  \item \EbnfNonTerminal{Attribute}: either \begin{inparaenum}[1)]
      \item Greek letter $@$,
      \item Greek letter $^$,
      or
      \item a string of lowercase English letters possibly with dashes inside, e.g. \enquote{\ff{price}} or \enquote{\ff{a-car}};
  \end{inparaenum}
  \item \EbnfNonTerminal{Data}: a sequence of bytes in hexadecimal format, e.g. \enquote{\ff{EF-41-5C}} is a sequence of three bytes, \enquote{\ff{42-}} is a one-byte sequence (with a trailing dash to avoid confusion with integers), and \enquote{\ff{-{}-}} (double dash) is an empty sequence of bytes;
  \item \EbnfNonTerminal{Function}: a string of English letters and numbers where the first symbol is an uppercase letter, e.g. \enquote{\ff{Sqrt}} or \enquote{\ff{F1}}.
\end{itemize}

\Cref{tab:sugar} shows all possible syntax sugar.

\begin{table*}
\capt{Syntax sugar.}
\label{tab:sugar}
\newcommand\sugar[2]{$ #1 $ & $ #2 $ \\}
\newcommand\subs[1]{& \textcolor{gray}{(#1)} \\}
\small
\begin{tabular}{ll}
\toprule
Syntax sugar & Its more verbose canonical equivalent \\
\midrule
\sugar
  {e ( \tau_1 -> e_1 , \tau_2 -> e_2 , \dots )}
  {e ( \tau_1 -> e_1 )( \tau_2 -> e_2 ) \dots}
\sugar
  {\tau_1 ( \tau_2 ,\; \tau_3 , \dots ) -> [[ B ]]}
  {\tau_1 -> [[ \tau_2 -> ? , \tau_3 -> ? , \dots , B ]]}
\sugar
  {\tau_1 -> \tau_2}
  {\tau_1 -> \phiTerminal{\upxi}.\tau_2}
\sugar
  {e:\tau}
  {[[ \tau -> e ]]}
\sugar
  {\texttt{\begin{CJK}{UTF8}{gbsn}"你好"\end{CJK}}}
  { Q.|string| ( @ -> Q.|bytes| ( @ -> [[ D> |E4-BD-A0-E5-A5-BD| ]] ) )}
  \subs{UTF-8 string}
\sugar
  {\texttt{42}}
  { Q.|number| ( @ -> Q.|bytes| ( @ -> [[ D> |40-45-00-00-00-00-00-00| ]] ) )}
  \subs{eight bytes per integer, assuming it has a floating point}
\sugar
  {\texttt{3.14}}
  { Q.|number| ( @ -> Q.|bytes| ( @ -> [[ D> |40-09-1E-B8-51-EB-85-1F| ]] ) )}
  \subs{eight bytes per number with a floating point}
\sugar
  { Q.|nan| }
  { Q.|number| ( @ -> Q.|bytes| ( @ -> [[ D> |7F-F8-00-00-00-00-00-00| ]] ) )}
\sugar
  { Q.|pinf| }
  { Q.|number| ( @ -> Q.|bytes| ( @ -> [[ D> |7F-F0-00-00-00-00-00-00| ]] ) )}
  \subs{positive infinity}
\sugar
  { Q.|ninf| }
  { Q.|number| ( @ -> Q.|bytes| ( @ -> [[ D> |FF-F0-00-00-00-00-00-00| ]] ) )}
  \subs{negative infinity}
\bottomrule
\end{tabular}
\end{table*}

% SPDX-FileCopyrightText: Copyright (c) 2016-2026 Objectionary.com
% SPDX-License-Identifier: MIT

\section{Foundations}\label{sec:foundations}

In this section, we introduce the core concepts of \phic{}.
The purpose of this section is to define the entities
  that constitute the calculus---such as
  expressions, formations, and attributes---together
  with their structural relationships.
The definitions given here are descriptive rather than operational.
They explain what expressions are, not how they are evaluated or transformed.

\begin{definition}[Expression]
An \textbf{expression}, ranged over \(\mathcal{E}\) by \(e_i\),
  is a grammatical construct obeying the syntax of \cref{fig:ebnf}.
\end{definition}

\begin{definition}[Attribute]
An \textbf{attribute}, ranged over \(\mathcal{T}\) by \(\tau_i\),
  is an identifier to which either $?$ or an expression
  may be attached.
\end{definition}

\begin{definition}[Void vs. Attached]
An attribute is \textbf{void} if $?$ is attached to it,
  otherwise it is an \textbf{attached} attribute.
\end{definition}

\begin{definition}[Datum]
A \textbf{datum}, ranged over \(\mathcal{D}\) by \(\delta_i\),
  is a possibly empty sequence of 8-bit bytes.
\end{definition}

\begin{definition}[State]
A \textbf{state} of evaluation \(s_i\) ranges over \(\mathcal{S}\) and
  represents the external world---such as the store and
  input/output---that a function reads and shifts from one state to the
  next when applied, its internal structure left unspecified.
\end{definition}

\begin{definition}[Universe]
A \textbf{universe} is an expression that serves as the global context of
  another expression, in which the $Q$ locator refers to the universe.
\end{definition}

\begin{definition}[Function]
A \textbf{function} is a total mapping
  \(\mathcal{B} \times \mathcal{E} \times \mathcal{S}
  \to \mathcal{E} \times \mathcal{S}\)
  that deterministically maps formation, universe,
  and state to an expression and a new state,
  ranged over \(\mathcal{F}\) by \(f_i\).
\end{definition}

The implementation of functions is outside the scope of \phic{}.

\begin{definition}[Asset]
An \textbf{asset} is an identifier to which either data
  (denoted as $D$-asset) or function
  (denoted as $L$-asset) is attached.
\end{definition}

\begin{definition}[Binding]
A \textbf{binding}, ranged over \(\mathcal{G}\) by \(B_i\),
is a possibly empty finite set of key-value pairs, where all keys are unique.
\end{definition}

The order in which the pairs of a binding are written is insignificant:
  $[[ |a| -> e_1, |b| -> e_2 ]]$ and $[[ |b| -> e_2, |a| -> e_1 ]]$
  denote the same formation.

\begin{definition}[Formation]
An object \textbf{formation}, denoted as $[[ B ]]$,
  ranged over \(\mathcal{B}\) by \(b_i\), is a binding.
\end{definition}

If $ b = [[ B ]] $, then the predicate \(k \in b\)
  holds if the key \(k\) is present in \(B\).

We introduce the term \enquote{formation} rather than using
  the more traditional term \enquote{construction}
  because the latter generally implies the presence of a
  class from which an object is being constructed or instantiated.
Instead, formation is closer to the creation of a prototype,
  which may either be used \enquote{as is} or copied.

The following is an example of a formation with four pairs, where the first one
  is an asset attached to data, while the other three are attributes attached to
  expressions:
\begin{phiquation}
\label{eq:object-example}
 [[ D> 00-2A, |b| -> b_2(|x| -> b_3).bar, |a| -> [[ L> |Sqrt| ]], foo -> T ]] {.}
\end{phiquation}

The arrow $ -> $ denotes an attachment of an expression (right-hand side)
  to an attribute (left-hand side).
The arrow $ ..> $ denotes an attachment of data or function to an asset.

\begin{definition}[Terminator]
The \textbf{terminator}, denoted as $T$,
  is an expression that signals an error.
\end{definition}

\begin{definition}[Abstract Formation]
A formation with at least one void attribute is an \textbf{abstract formation}.
\end{definition}

\Cref{eq:price-color} is an example of an abstract formation,
  while the formation attached to its \ff{color} attribute is closed.

A formation that is not an abstract formation may be called a \emph{closed} formation.

\begin{definition}[Application]
An \textbf{application},
  denoted as $ e_1 ( \tau -> e_2 ) $,
  means an attempt to attach \(e_2\) (the \enquote{argument}) to
  the \(\tau\) attribute of \(e_1\) (the \enquote{subject}).
\end{definition}

\Cref{eq:simple-application} demonstrates application,
  where $ |x| -> b $ is applied to an abstract formation $[[ |x| -> ? ]]$.
The application creates a new formation $[[ |x| -> b ]]$,
  while the existing formation remains intact.

The formation in \cref{eq:price-color} is an abstract formation,
  because its attribute \ff{price} is void.
The formation in \cref{eq:simple-application}
  was an abstract formation before the application,
  but the formation created by the application is closed
  since its attribute \ff{x} is not void (attached to \(b\)).

Even though $?$ being attached to an attribute of a formation
  resembles NULL references, there is a significant difference:
  in \phic{}, void attributes may be attached to expressions only once,
  while any further reattachments are prohibited.

\begin{definition}[Dispatch]
A \textbf{dispatch},
  denoted as $ e . \tau $, where \(e\) is the \enquote{subject,}
  means an attempt to retrieve what is attached to \(\tau\)
  in the formation to which \(e\) may be transformed.
\end{definition}

\begin{definition}[Atom]
An \textbf{atom} is a formation with a function attached to its $L$-asset.
\end{definition}

\Cref{eq:Sqrt} demonstrates an atom with a function
  that calculates the square root of a number,
  which it retrieves from the $D$-asset
  of $b.|x|$ with the help of the morphing function (\cref{sec:morphing}).

\begin{definition}[Decorator]
A formation is a \textbf{decorator} if it has
  an $@$-attribute with an expression attached to it;
  the expression attached to the $@$-attribute is its \textbf{decoratee}.
\end{definition}

\begin{definition}[Parent]
Attaching expression \(e\) to the $^$ attribute of formation \(b\)
  means setting the \textbf{parent} of \(b\) to \(e\).
\end{definition}

The presence of \enquote{parent} in each formation is essential
  for the coordination of inner formations after dispatch.
Consider the following abstract formation with an inner formation:
\begin{phiquation*}
[[ |x| -> [[ |a| -> ?, next -> [[ @ -> $.^.|a|.plus( |x| -> 1 ) ]] ]], |k| -> $.|x|( |a| -> |42| ) ]] {.}
\end{phiquation*}
Here, if the parent of $Q.|x|.|next|$ were attached in the formation,
  the result of dataization of $Q.|k|.|next|$
  would not be equal to \ff{43}.
Instead, it would be equal to $T$, because $ Q.|k|.|next|.^ $
  would still be attached to $ Q.|x| $ after the dispatch of $ Q.|k|.next $.
The parent attribute may be compared with the \ff{this} pointer in Java or C++,
  which does not point anywhere until a method of a class is called.
Then, when the method is called, the \ff{this} pointer
  refers to the formation that owns the method.

% SPDX-FileCopyrightText: Copyright (c) 2016-2026 Objectionary.com
% SPDX-License-Identifier: MIT

\section{Operators}\label{sec:operators}

In this section we introduce a family of semantic operators
  defined on \(\upvarphi\)-expressions.
These operators do not extend the syntax of the calculus.
Instead, they describe systematic transformations of expressions that reveal,
  modify, or project their semantic content.
Each operator is defined in terms of the reduction semantics introduced earlier
  and preserves the core meaning of expressions
  while possibly changing their form or role.

\subsection{Contextualization}\label{sec:contextualization}
\makeatletter
\protected@edef\@currentlabelname{Ctx}%
\makeatother
\label{r:contextualize}

\begin{definition}[Absolute]
An expression is \textbf{absolute},
  denoted by \(k\) and ranging over \(\mathcal{K} \subseteq \mathcal{E}\),
  if it is either
  \begin{inparaenum}[a)]
  \item $Q$,
  \item formation,
  \item dispatch with an absolute subject,
  or
  \item application with absolute subject and argument.
  \end{inparaenum}
\end{definition}

\begin{definition}[Contextualization]
The \emph{contextualization} function
  \(\mathcal{E} \times \mathcal{K} \to \mathcal{E}\),
  denoted as \(\phinoContextualize{e_1}{k}{e_2}\)
  and specified by the rules of \cref{fig:contextualization},
  turns an expression \(e_1\) into \(e_2\)
  by replacing with the absolute \(k\)
  every \(\phiTerminal{\upxi}\) that denotes the context of \(e_1\)
  rather than that of a nested formation.
\end{definition}

\begin{figure*}
\begin{mdframed}
\centering
\iexec[maybe]{phino --pin-file=phino-version.txt explain --contextualize > \phinotex/rules-contextualize.tex}
\input{\phinotex/rules-contextualize.tex}
\end{mdframed}
\capt{Contextualization by induction, generated by phino \iexec[maybe]{phino --version}\unskip.}
\label{fig:contextualization}
\end{figure*}

\begin{theorem}\label{thm:contextualization-total}
The contextualization function is total, meaning that
  every expression can be contextualized and its derivation terminates.
\end{theorem}

\subsection{Normalization}\label{sec:normalization}

An expression that may be rewritten by the \emph{rules} (or \emph{reductions})
  listed in \cref{fig:reduction} is a \emph{reducible} expression.
The notation \(\phinoNormalize{e_1}{e_2}\), optionally followed by a condition,
  denotes a reduction of \(e_1\) to \(e_2\), if the condition holds.

A specific reduction may be denoted, for example,
  as \(\phiNormalize[\nameref{r:dot}]\),
  or just \(\phiNormalize\) when no specific reduction is meant.
The notation \(e_1 \phiTransNormalize e_2\) denotes a reflexive transitive
  closure of all reductions, so that there is a possibly empty finite
  sequence of reductions between \(e_1\) and \(e_2\).

\begin{definition}[Normal Form]
An expression that has no more possible applications of reductions
  is \emph{irreducible} or a \emph{normal form}, denoted as \(\nf{}_i\)
  ranging over \(\mathcal{N} \subseteq \mathcal{E}\).
\end{definition}

Thus, \(\nf\) is a normal form of \(e\) if \(e \phiTransNormalize \nf\) and
  there is no expression \(e_1\) such that \(\phinoNormalize{\nf}{e_1}\).

\begin{figure*}
\begin{mdframed}
\renewcommand{\arraystretch}{1.2}
\iexec[maybe]{phino --pin-file=phino-version.txt explain --normalize > \phinotex/rules.tex}

  \adjustbox{max width=\linewidth}{
  \begin{tabular}{rl}
  \input{\phinotex/rules.tex}
  \end{tabular}}
\end{mdframed}
\capt{Reduction rules, generated by phino \iexec[maybe]{phino --version}\unskip.}
\label{fig:reduction}
\end{figure*}

\begin{theorem}\label{thm:normalization-total}
Normalization is \emph{total}:
  every expression has a normal form and its reduction terminates.
\end{theorem}

\Cref{app:normalization-examples} demonstrates how normalization
  works through examples.

\begin{theorem}\label{thm:normalization-confluent}
The reduction relation is \emph{confluent}:
  if \(e \phiTransNormalize e_1\) and \(e \phiTransNormalize e_2\),
  then there exists an expression \(e_3\) such that
  \(e_1 \phiTransNormalize e_3\) and \(e_2 \phiTransNormalize e_3\).
\end{theorem}

Confluence guarantees that the normal form of an expression is unique,
  irrespective of the order in which the rules are applied.

We use Lean4 to automate the proof of \cref{thm:normalization-confluent},
  in the
  \href{https://github.com/objectionary/proof}{objectionary/proof}
  GitHub repository.

\subsection{Equivalence}\label{sec:equivalence}

\begin{definition}[Equivalence]
Two expressions are said to be syntactically equivalent or just
  \textbf{equivalent} (denoted by \(\equiv\)) if their normal forms are
  syntactically identical.
\end{definition}

\subsection{Evaluation}\label{sec:evaluation}
\makeatletter
\protected@edef\@currentlabelname{Evt}%
\makeatother
\label{r:evaluate}

\begin{definition}[Evaluation]
The \emph{evaluation} function
  \(\mathcal{B} \times \mathcal{E} \times \mathcal{S}
    \rightharpoonup \mathcal{N} \times \mathcal{S} \),
  denoted as \(\phinoEvaluate{b}{e}{s_1}{n}{s_2}\)
  and specified by the rules of \cref{fig:evaluation},
  maps an atom \(b\) to \(n\),
  calling by value the function attached to its $L$-asset
  in the universe \(e\),
  possibly shifting the state \(s_1\) to \(s_2\),
\end{definition}

\begin{figure*}
\begin{mdframed}
\begin{gather*}
\begin{phinoInference}
  \phinoCondition{ \langle e_1 {,}\; s_2 \rangle = f \lparen [[ B_1, L>f, B_2 ]] {,}\; e {,}\; s_1 \rparen }
  \phinoPremise{ \phinoNormalize{e_1}{n} }
  \phinoConclusion{ \phinoEvaluate{[[ B_1, L>f, B_2 ]]}{e}{s_1}{n}{s_2} }
\end{phinoInference}
\\
\begin{phinoInference}
  \phinoCondition{ \uplambda \not\in B }
  \phinoConclusion{ \phinoEvaluate{[[ B ]]}{e}{s}{ [[ B ]]}{s} }
\end{phinoInference}
\end{gather*}
\end{mdframed}
\capt{Evaluation rules.}
\label{fig:evaluation}
\end{figure*}

Because the function attached to an atom is arbitrary---its implementation
  lies outside the calculus (\cref{sec:foundations})---firing it may run any
  computation, and the calculus cannot guarantee that a datum ever comes back.
Evaluation is therefore partial on two counts: the call may diverge, as when an
  atom dispatches back into its own object and recurs without bound, or it may
  halt without a value, as when the function aborts on an undefined operation
  such as division by zero.
Morphing and dataization reach atoms and inherit that partiality,
  whereas contextualization and normalization fire no atom and are total
  (\cref{thm:contextualization-total,thm:normalization-total}).

\subsection{Morphing}\label{sec:morphing}

\begin{definition}[Morphing]
The \emph{morphing} function
  \(\mathcal{N} \times \mathcal{E} \times \mathcal{S}
    \rightharpoonup ( \mathcal{B} \cup \{ \phiTerminal{\bot} \} ) \times \mathcal{S} \),
  denoted as \(\phinoMorph{n_1}{e}{s_1}{n_2}{s_2}\)
  and specified by the rules of \cref{fig:morphing},
  reduces a normal form \(n_1\) to the formation it denotes,
  or to $T$ if it denotes none,
  by evaluating the atoms it reaches
  and resolving the $Q$ locator against the universe \(e\),
  possibly shifting the state \(s_1\) to \(s_2\).
\end{definition}

Morphing is a partial function, since the atoms it reaches
  (\cref{sec:evaluation}) may fire without bound,
  and it may follow a cyclic decoration chain that never resolves.

\begin{figure*}
\begin{mdframed}
\centering
\iexec[maybe]{phino --pin-file=phino-version.txt explain --morph > \phinotex/rules-morph.tex}
\input{\phinotex/rules-morph.tex}
\end{mdframed}
\capt{Morphing rules, generated by phino \iexec[maybe]{phino --version}\unskip.}
\label{fig:morphing}
\end{figure*}

\begin{theorem}\label{thm:morphing-confluent}
The morphing function is confluent, meaning that
  every normal form can be morphed to at most one formation or $T$.
\end{theorem}

\subsection{Dataization}\label{sec:dataization}

\begin{definition}[Dataization]
The \emph{dataization} function
  \(\mathcal{N} \times \mathcal{E} \times \mathcal{S}
    \rightharpoonup \mathcal{D} \times \mathcal{S}\),
  denoted as \(\phinoDataize{n}{e}{s_1}{\delta}{s_2}\)
  and specified by the rules of \cref{fig:dataization},
  extracts the datum \(\delta\) that a normal form \(n\) produces
  by morphing \(n\) to a formation in the universe \(e\)
  and reading the data attached to its $D$-asset,
  possibly shifting the state \(s_1\) to \(s_2\).
\end{definition}

\begin{theorem}[Progress]\label{thm:dataization-progress}
At least one rule of \cref{fig:dataization} applies to every normal form except
  $T$, which matches none, so a dataization derivation halts only at a datum or
  at $T$, the error that $T$ signals.
\end{theorem}

\begin{figure*}
\begin{mdframed}
\centering
\iexec[maybe]{phino --pin-file=phino-version.txt explain --dataize > \phinotex/rules-dataize.tex}
\input{\phinotex/rules-dataize.tex}
\end{mdframed}
\capt{Dataization rules, generated by phino \iexec[maybe]{phino --version}\unskip.}
\label{fig:dataization}
\end{figure*}

\begin{theorem}\label{thm:dataization-confluent}
The dataization function is confluent, meaning that
  every normal form can be dataized to at most one datum.
\end{theorem}

\Cref{app:dataization-examples} demonstrates how the dataization function works through examples.

\subsection{Congruence}

\begin{definition}[Congruence]
Two expressions \(n_1\) and \(n_2\) are said to be behaviorally equivalent
  or \textbf{congruent} (denoted by \(\cong\))
  if for any state \(s_1\) and any \(e\) their dataizations are either both
  undefined, or both defined with the same datum and state,
  i.e. there exist \(\delta\) and \(s_2\) such that:
\begin{phiquation*}
\phinoDataize{n_1}{e}{s_1}{\delta}{s_2} \quad\text{and}\quad \phinoDataize{n_2}{e}{s_1}{\delta}{s_2}{.}
\end{phiquation*}
\end{definition}

Dataization gives the calculus its intended, observational semantics: the
  meaning of an expression is the datum it computes under a given universe and
  state, and two objects denote the same behavior exactly when they are
  congruent.
The reduction rules are sound with respect to this meaning, since equivalent
  expressions are congruent (\(\equiv\) implies \(\cong\))---identical normal
  forms dataize identically---so normalizing an expression never changes what
  it computes.

Two congruent expressions may be non-equivalent.
For example, \(e_1 \cong e_2\) holds while \(e_1 \equiv e_2\) doesn't hold
  for the following expressions:
\begin{phiquation*}
e_1 = [[ foo -> ?, D> 01-02 ]] \quad e_2 = [[ bar -> ?, D> 01-02 ]]{.}
\end{phiquation*}

% SPDX-FileCopyrightText: Copyright (c) 2016-2026 Objectionary.com
% SPDX-License-Identifier: MIT

\section{Related Work}\label{sec:related}

In this section we analyze and categorize prior art related to our work.
Neither object-oriented formalism nor pure object-oriented languages
  are new research topics.
However, we identified certain gaps in existing studies
  that make us believe that our work has novelty.

Attempts were made to formalize OOP and introduce an object calculus
  similar to the lambda calculus~\citep{barendregt2012}
  used in functional programming.
For example, \citet{abadi1995imperative} suggested an imperative
  calculus of objects, which was extended by~\citet{bono1998imperative}
  to support classes, by~\citet{gordon1998concurrent} to support concurrency
  and synchronization, and by~\citet{jeffrey1999distributed}
  to support distributed programming.
These calculi, together with their functional counterpart,
  were consolidated into a comprehensive
  theory of objects~\citep{abadi1996theory}.

A parallel line of work embeds objects directly into the
  \(\lambda\)-calculus rather than defining a standalone calculus.
\citet{mitchell1993lambda} introduced a lambda calculus of objects
  with method specialization, extended by~\citet{di1998lambda}
  with self-inflicted extension.
These systems show that object features can be encoded on a functional core,
  which qualifies the recurring claim that OOP lacks
  a \(\lambda\)-calculus of its own; \phic{} takes the opposite route,
  treating the object rather than the function as its single primitive,
  so that dispatch and decoration are native constructs rather than encodings.

Earlier, \citet{honda1991object} combined OOP and \(\pi\)-calculus in order
  to introduce object calculus for asynchronous communication,
  which was further referenced by~\citet{jones1993pi} in their work
  on object-based design notation.

A few attempts were made to reduce existing OOP languages
  and formalize what is left.
Featherweight Java is the most notable example proposed
  by~\citet{igarashi2001featherweight}, which omits almost all features
  of the full language (including interfaces and even assignment)
  to obtain a small calculus.
Later it was extended by~\citet{jagannathan2005transactional}
  to support nested and multi-threaded transactions.
Featherweight Java is used in formal languages such as
  Obsidian~\citep{coblenz2019} and SJF~\citep{usov2020}.

The calculi above are class-based, whereas \phic{} is prototype-based:
  objects arise by copying and decorating other objects rather
  than by instantiating classes.
This tradition begins with Self~\citep{ungar1987self},
  whose delegation mechanism \citet{stein1987delegation} showed to subsume
  inheritance, and continues in calculi of delegation
  by object composition~\citep{bettinibono2008delegation}.
The meaning of inheritance in such settings has been given
  denotationally by~\citet{cook1989denotational} as a fixed point over records;
  \phic{} instead renders it operationally, through the $@$-decoration that
  its morphing operator resolves (\cref{sec:morphing}).

Another example is Larch/C++~\citep{cheon1994quick},
  which is a formal algebraic interface specification language tailored to C++.
It allows interfaces of C++ classes and functions to be documented
  in a way that is unambiguous and concise.

Several attempts to formalize OOP were made by extensions of
  the most popular formal notations and methods,
  such as Object-Z~\citep{duke1991object} and VDM++~\citep{durr1992vdm}.
In Object-Z, state and operation schemes are encapsulated into classes.
The formal model is based upon the idea
  of a class history~\citep{duke1990towards}.
However, all these OO extensions do not have comprehensive refinement
  rules that can be used to transform specifications into implemented code
  in an actual OO programming language, as was noted by~\citet{paige1999object}.

\citet{bancilhon1985calculus} suggested an object calculus
  as an extension to relational calculus.
\citet{jankowska2003anotheroop} further developed these ideas and related
  them to a Boolean algebra.
\citet{leekwakryu1996transform} developed an algorithm to transform
  an object calculus into an object algebra.

However, all these theoretical attempts to formalize OO languages were unable
  to fully describe their features,
  as was noted by~\citet{nierstrasz1991towards}: ``The development of
  concurrent object-based programming languages has suffered from
  the lack of any generally accepted formal foundations
  for defining their semantics.''
The visual and architectural notations often used to model
  object-oriented designs fare worse still:
  \citet{eden2002visual}, surveying them, observed that
  ``Not one of the notations is defined formally,
  nor provided with \nospell{denotational} semantics,
  nor founded on axiomatic semantics.''
Moreover, \citet{ciaffaglione2003reasoning,ciaffaglione2003typetheories,ciaffaglione2007theory_of_contexts}
  noted in their series of works that, at the time,
  relatively little formal work had been carried out on object-based languages.
Mechanized treatments have appeared since---among them
  a machine-checked model of a Java-like language,
  virtual machine, and compiler~\citep{klein2006jinja}---yet a compact
  object calculus whose own meta-theory is machine-checked remains uncommon.

What sets \phic{} apart from the surveyed formalisms is therefore
  not broader feature coverage, which it deliberately forgoes
  (\cref{sec:future}), but this uniformity together with
  a confluence result certified in Lean~4 (\cref{thm:normalization-confluent}).

% SPDX-FileCopyrightText: Copyright (c) 2016-2026 Objectionary.com
% SPDX-License-Identifier: MIT

\section{Future Studies}\label{sec:future}

This paper intentionally focuses on defining \phic{} (syntax, foundations, and semantic operators) and does not yet demonstrate practical use.
At the same time, it motivates the calculus as a step toward more rigorous reasoning,
  verified compilers and static analyzers, and better language design.
Future work may build on \phic{} by creating the following artifacts:

\begin{itemize}
  \item A short list of core laws about reductions (what always works, what may fail).
  \item A clear way to prove that two expressions behave the same in any context.
  \item A type system that catches mistakes early.
  \item A way to mark which expressions are pure and which may have side effects.
  \item A catalog of safe rewrites (how to change code without changing meaning).
  \item A translation from Java or C++ into \(\upvarphi\)-expressions.
  \item A standard library of built-in atoms with precise rules.
  \item Larger language features built as add-ons (overriding, interfaces, constructors).
  \item Add-ons for mutation, concurrency, and distribution, with simple safety guarantees.
\end{itemize}

We invite readers to extend, formalize, and apply \phic{} to real object-oriented patterns and languages,
  and to help turn these foundations into a shared body of results and artifacts.

% SPDX-FileCopyrightText: Copyright (c) 2016-2026 Objectionary.com
% SPDX-License-Identifier: MIT

\section{Acknowledgments}

Many thanks to (in alphabetical order of last names)
  \nospell{Aliaksei Bial\'{i}auski},
  \nospell{Fabricio Cabral},
  \nospell{Kirill Chernyavskiy},
  \nospell{Piotr Chmielowski},
  \nospell{Danilo Danko},
  \nospell{Konstantin Gukov},
  \nospell{Andrew Graur},
  \nospell{Ali-Sultan Ki\-giz\-ba\-ev},
  \nospell{Nikolai Ku\-da\-sov},
  \nospell{Alexander Legalov},
  \nospell{Mikhail Lipanin},
  \nospell{Tymur \(\lambda\)ysenko},
  \nospell{Alexandr Naumchev},
  \nospell{Alonso A. Ortega},
  \nospell{John Page},
  \nospell{Alex Panov},
  \nospell{Maxim Petrov},
  \nospell{Alexander Pushkarev},
  \nospell{Marcos Douglas B. Santos},
  \nospell{Alex Semenyuk},
  \nospell{Violetta Sim},
  \nospell{Sergei Skliar},
  \nospell{Stian Soiland-Reyes},
  \nospell{Viacheslav Tradunskyi},
  \nospell{Ilya Trub},
  \nospell{César Soto Valero},
  \nospell{Alena Vasileva},
  \nospell{David West},
  \nospell{Vladimir Zakharov},
  and
  \nospell{Evgeny Zouev}
for their contribution to the development of \phic{}.

{\raggedright
\bibliographystyle{ACM-Reference-Format}
\begin{multicols}{2}
\bibliography{bibliography/main}
\end{multicols}}
\vfill\eject

\appendix

% SPDX-FileCopyrightText: Copyright (c) 2016-2026 Objectionary.com
% SPDX-License-Identifier: MIT

\newpage
\section{Examples of Normalization}
\label{app:normalization-examples}

\newcounter{exmp}
\newcommand\phiExpression[1]{%
  \stepcounter{exmp}
  e_{\theexmp} = }

The following examples demonstrate how the reduction rules of \cref{fig:reduction} may normalize
some expressions, involving the contextualization function (\cref{sec:contextualization}):

\iexec[maybe]{./examples-to-tex.sh examples/basic > \phinotex/examples.tex}
\begin{phiquation*}
\phinoMeet{basic-a:1}{ [[ |x| -> |t|, |t| -> ? ]] } . |x| \phiNormalize[\nameref{r:dot}]
  \phiNormalize ? : |t| . |t| ( \phiTerminal{\rho} -> \phinoAgain{basic-a:1} ) \phiNormalize[\nameref{r:null}]
  \phiNormalize T ( \phiTerminal{\rho} -> \phinoAgain{basic-a:1} ) \phiNormalize[\nameref{r:dc}]
  \phiNormalize T{.}
\end{phiquation*}
\begin{phiquation*}
\phinoMeet{basic-b:1}{ |42.0| : |t| } . |t| : |x| . |x| \phiNormalize[\nameref{r:dot}]
  \phiNormalize |42.0| : |x| . |x| \phiNormalize[\nameref{r:dot}]
  \phiNormalize |42.0|{.}
\end{phiquation*}
\begin{phiquation*}
[[ |x| -> |Fn| : L . \phiTerminal{\rho} . |k|, |k| -> |42.0| ]] . |x| \phiNormalize[\nameref{r:norho}]
  \phiNormalize [[ |x| -> T . |k|, |k| -> |42.0| ]] . |x| \phiNormalize[\nameref{r:dd}]
  \phiNormalize \phinoMeet{basic-e:1}{ [[ |x| -> T, |k| -> |42.0| ]] } . |x| \phiNormalize[\nameref{r:dot}]
  \phiNormalize T ( \phiTerminal{\rho} -> \phinoAgain{basic-e:1} ) \phiNormalize[\nameref{r:dc}]
  \phiNormalize T{.}
\end{phiquation*}
\begin{phiquation*}
[[ |x| -> ?, |y| -> |x| ]] ( |x| -> |42.0| ) . |y| \phiNormalize[\nameref{r:copy}]
  \phiNormalize \phinoMeet{basic-f:1}{ [[ |x| -> |42.0|, |y| -> |x| ]] } . |y| \phiNormalize[\nameref{r:dot}]
  \phiNormalize |42.0| : |x| . |x| ( \phiTerminal{\rho} -> \phinoAgain{basic-f:1} ) \phiNormalize[\nameref{r:dot}]
  \phiNormalize |42.0|{.}
\end{phiquation*}
\begin{phiquation*}
[[ |x| -> |t|, |t| -> ? ]] ( |t| -> |42.0| ) \phiNormalize[\nameref{r:copy}]
  \phiNormalize [[ |x| -> |t|, |t| -> |42.0| ]]{.}
\end{phiquation*}
\begin{phiquation*}
[[ |x| -> |t|, @ -> [[]] : |t| ]] . |x| \phiNormalize[\nameref{r:dot}]
  \phiNormalize [[]] : |t| : @ . |t| ( \phiTerminal{\rho} -> [[ |x| -> |t|, @ -> [[]] : |t| ]] ){.}
\end{phiquation*}
\begin{phiquation*}
\phinoMeet{basic-i:1}{ \phiTerminal{\rho} . \phiTerminal{\rho} . |t| : |x| } . |x| \phiNormalize[\nameref{r:dot}]
  \phiNormalize [[]] . \phiTerminal{\rho} . \phiTerminal{\rho} . |t| ( \phiTerminal{\rho} -> \phinoAgain{basic-i:1} ) \phiNormalize[\nameref{r:norho}]
  \phiNormalize T . \phiTerminal{\rho} . |t| ( \phiTerminal{\rho} -> \phinoAgain{basic-i:1} ) \phiNormalize[\nameref{r:dd}]
  \phiNormalize T . |t| ( \phiTerminal{\rho} -> \phinoAgain{basic-i:1} ) \phiNormalize[\nameref{r:dd}]
  \phiNormalize T ( \phiTerminal{\rho} -> \phinoAgain{basic-i:1} ) \phiNormalize[\nameref{r:dc}]
  \phiNormalize T{.}
\end{phiquation*}
\begin{phiquation*}
[[ |x| -> [[ \phiTerminal{\rho} -> ? ]] . \phiTerminal{\rho} . |k|, |k| -> |42.0| ]] . |x| \phiNormalize[\nameref{r:null}]
  \phiNormalize [[ |x| -> T . |k|, |k| -> |42.0| ]] . |x| \phiNormalize[\nameref{r:dd}]
  \phiNormalize \phinoMeet{basic-j:1}{ [[ |x| -> T, |k| -> |42.0| ]] } . |x| \phiNormalize[\nameref{r:dot}]
  \phiNormalize T ( \phiTerminal{\rho} -> \phinoAgain{basic-j:1} ) \phiNormalize[\nameref{r:dc}]
  \phiNormalize T{.}
\end{phiquation*}
\begin{phiquation*}
\phinoMeet{basic-k:1}{ [[ |x| -> |t| ( |k| -> [[]] ) . |k|, |t|(|k|) -> [[]] ]] } . |x| \phiNormalize[\nameref{r:dot}]
  \phiNormalize [[ |t|(|k|) -> [[]] ]] . |t| ( |k| -> [[]] ) . |k| ( \phiTerminal{\rho} -> \phinoAgain{basic-k:1} ) \phiNormalize[\nameref{r:dot}]
  \phiNormalize ? : |k| ( \phiTerminal{\rho} -> [[ |t|(|k|) -> [[]] ]], |k| -> [[]] ) . |k| ( \phiTerminal{\rho} -> \phinoAgain{basic-k:1} ) \phiNormalize[\nameref{r:skip}]
  \phiNormalize ? : |k| ( |k| -> [[]] ) . |k| ( \phiTerminal{\rho} -> \phinoAgain{basic-k:1} ) \phiNormalize[\nameref{r:copy}]
  \phiNormalize [[]] : |k| . |k| ( \phiTerminal{\rho} -> \phinoAgain{basic-k:1} ) \phiNormalize[\nameref{r:dot}]
  \phiNormalize [[]] ( \phiTerminal{\rho} -> [[]] : |k|, \phiTerminal{\rho} -> \phinoAgain{basic-k:1} ) \phiNormalize[\nameref{r:skip}]
  \phiNormalize [[]] ( \phiTerminal{\rho} -> \phinoAgain{basic-k:1} ) \phiNormalize[\nameref{r:skip}]
  \phiNormalize [[]]{.}
\end{phiquation*}
\begin{phiquation*}
\phinoMeet{basic-l:1}{ [[ |x| -> ? : |t| ( |t| -> |k| ), |k| -> |Fn| : L ]] } . |x| . |t| . |p| : |a| \phiNormalize[\nameref{r:dot}]
  \phiNormalize ? : |t| ( |t| -> |Fn| : L : |k| . |k|, \phiTerminal{\rho} -> \phinoAgain{basic-l:1} ) . |t| . |p| : |a| \phiNormalize[\nameref{r:dot}]
  \phiNormalize ? : |t| ( |t| -> |Fn| : L ( \phiTerminal{\rho} -> |Fn| : L : |k| ), \phiTerminal{\rho} -> \phinoAgain{basic-l:1} ) . |t| . |p| : |a| \phiNormalize[\nameref{r:skip}]
  \phiNormalize ? : |t| ( |t| -> |Fn| : L, \phiTerminal{\rho} -> \phinoAgain{basic-l:1} ) . |t| . |p| : |a| \phiNormalize[\nameref{r:copy}]
  \phiNormalize \phinoMeet{basic-l:2}{ |Fn| : L : |t| } ( \phiTerminal{\rho} -> \phinoAgain{basic-l:1} ) . |t| . |p| : |a| \phiNormalize[\nameref{r:skip}]
  \phiNormalize \phinoAgain{basic-l:2} . |t| . |p| : |a| \phiNormalize[\nameref{r:dot}]
  \phiNormalize |Fn| : L ( \phiTerminal{\rho} -> \phinoAgain{basic-l:2} ) . |p| : |a| \phiNormalize[\nameref{r:skip}]
  \phiNormalize |Fn| : L . |p| : |a|{.}
\end{phiquation*}
\begin{phiquation*}
[[ |x| -> \phiTerminal{\rho} . \phiTerminal{\rho} . |k| : |p| . |p| : |t| . |t|, |k| -> |42.0| ]] . |x| \phiNormalize[\nameref{r:dot}]
  \phiNormalize [[ |x| -> [[]] . \phiTerminal{\rho} . \phiTerminal{\rho} . |k| ( \phiTerminal{\rho} -> \phiTerminal{\rho} . \phiTerminal{\rho} . |k| : |p| ) : |t| . |t|, |k| -> |42.0| ]] . |x| \phiNormalize[\nameref{r:norho}]
  \phiNormalize [[ |x| -> T . \phiTerminal{\rho} . |k| ( \phiTerminal{\rho} -> \phiTerminal{\rho} . \phiTerminal{\rho} . |k| : |p| ) : |t| . |t|, |k| -> |42.0| ]] . |x| \phiNormalize[\nameref{r:dd}]
  \phiNormalize [[ |x| -> T . |k| ( \phiTerminal{\rho} -> \phiTerminal{\rho} . \phiTerminal{\rho} . |k| : |p| ) : |t| . |t|, |k| -> |42.0| ]] . |x| \phiNormalize[\nameref{r:dd}]
  \phiNormalize [[ |x| -> T ( \phiTerminal{\rho} -> \phiTerminal{\rho} . \phiTerminal{\rho} . |k| : |p| ) : |t| . |t|, |k| -> |42.0| ]] . |x| \phiNormalize[\nameref{r:dc}]
  \phiNormalize [[ |x| -> T : |t| . |t|, |k| -> |42.0| ]] . |x| \phiNormalize[\nameref{r:dot}]
  \phiNormalize [[ |x| -> T ( \phiTerminal{\rho} -> T : |t| ), |k| -> |42.0| ]] . |x| \phiNormalize[\nameref{r:dc}]
  \phiNormalize \phinoMeet{basic-m:1}{ [[ |x| -> T, |k| -> |42.0| ]] } . |x| \phiNormalize[\nameref{r:dot}]
  \phiNormalize T ( \phiTerminal{\rho} -> \phinoAgain{basic-m:1} ) \phiNormalize[\nameref{r:dc}]
  \phiNormalize T{.}
\end{phiquation*}
\begin{phiquation*}
[[ |x| -> \phinoMeet{basic-n:1}{ \phiTerminal{\rho} . |k| . \phiTerminal{\rho} . |t| : |t| }, |k| -> [[]], |t| -> [[]] ]] . |x| . |t| \phiNormalize[\nameref{r:dot}]
  \phiNormalize \phinoAgain{basic-n:1} ( \phiTerminal{\rho} -> [[ |x| -> \phinoAgain{basic-n:1}, |k| -> [[]], |t| -> [[]] ]] ) . |t| \phiNormalize[\nameref{r:skip}]
  \phiNormalize \phinoAgain{basic-n:1} . |t| \phiNormalize[\nameref{r:dot}]
  \phiNormalize [[]] . \phiTerminal{\rho} . |k| . \phiTerminal{\rho} . |t| ( \phiTerminal{\rho} -> \phinoAgain{basic-n:1} ) \phiNormalize[\nameref{r:norho}]
  \phiNormalize T . |k| . \phiTerminal{\rho} . |t| ( \phiTerminal{\rho} -> \phinoAgain{basic-n:1} ) \phiNormalize[\nameref{r:dd}]
  \phiNormalize T . \phiTerminal{\rho} . |t| ( \phiTerminal{\rho} -> \phinoAgain{basic-n:1} ) \phiNormalize[\nameref{r:dd}]
  \phiNormalize T . |t| ( \phiTerminal{\rho} -> \phinoAgain{basic-n:1} ) \phiNormalize[\nameref{r:dd}]
  \phiNormalize T ( \phiTerminal{\rho} -> \phinoAgain{basic-n:1} ) \phiNormalize[\nameref{r:dc}]
  \phiNormalize T{.}
\end{phiquation*}

The expressions at \cref{fig:normals} may not be reduced any further;
  they are in normal form.

\begin{figure*}
\begin{mdframed}
\begin{multicols}{2}
\iexec[maybe]{./examples-to-tex.sh examples/nf > \phinotex/examples-nf.tex}
\input{\phinotex/examples-nf.tex}
\end{multicols}
\end{mdframed}
\capt{Normal forms, checked by phino \iexec[maybe]{phino --version}\unskip.}
\label{fig:normals}
\end{figure*}

% SPDX-FileCopyrightText: Copyright (c) 2016-2026 Objectionary.com
% SPDX-License-Identifier: MIT

\newpage
\section{Example of Dataization}
\label{app:dataization-examples}

Consider an algorithm that converts degrees from Celsius to Fahrenheit:
\begin{equation*}
^{\circ}F ={} ^{\circ}C \times 1.8 + 32,
\end{equation*}
Assuming that \(^{\circ}C\) already equals \ff{25} and is attached
  to the \ff{c} attribute instead of being provided by a user,
  the algorithm may be represented by an expression:
\iexec[maybe]{phino --pin-file=phino-version.txt rewrite --output=latex --sweet --label=eq:celsius --margin=80 ./examples/celsius.phi > \phinotex/celsius-formula.tex}
\begin{phiquation}
\label{eq:celsius}
[[
  @ -> [[ @ -> |c| . |times| ( |x| -> |1.8| ) . |plus| ( |x| -> |32.0| ), |c| -> |25.0| ]],
  |bytes|(@) -> [[]],
  |number|(@) -> [[ |times|(|x|, \phiTerminal{\rho}) -> |Fn1| : L, |plus|(|x|, \phiTerminal{\rho}) -> |Fn2| : L ]]
]]{.}
\end{phiquation}

The evaluations of \texttt{Fn1} and \texttt{Fn2} are defined as follows:
\begin{align*}
\adjustbox{max width=\linewidth,margin=0em .3em}{%
  \text{\texttt{Fn1}:}\;%
  \ensuremath{\infer{\text{\phiq{\phinoEvaluate{b}{e}{s}{ n_3 }{s}}}}{%
    \infer[\;\text{if}\;\text{\phiq{\delta = \delta_1 \times \delta_2}}]{\text{\phiq{\phinoNormalize{ e.number( @ -> e.bytes( @ -> [[ D> \delta ]] ) ) }{ n_3 }}}}{%
      \infer{\text{\phiq{\phinoDataize{n_1}{e}{s}{\delta_1}{s}}}}{%
        \text{\phiq{\phinoNormalize{b.^}{n_1}}}%
      }%
      &
      \infer{\text{\phiq{\phinoDataize{n_2}{e}{s}{\delta_2}{s}}}}{%
        \text{\phiq{\phinoNormalize{b.|x|}{n_2}}}%
      }%
    }%
  }}%
}
\\
\adjustbox{max width=\linewidth,margin=0em .3em}{%
  \text{\texttt{Fn2}:}\;%
  \ensuremath{\infer{\text{\phiq{\phinoEvaluate{b}{e}{s}{ n_3 }{s}}}}{%
    \infer[\;\text{if}\;\text{\phiq{\delta = \delta_1 + \delta_2}}]{\text{\phiq{\phinoNormalize{ e.number( @ -> e.bytes( @ -> [[ D> \delta ]] ) ) }{ n_3 }}}}{%
      \infer{\text{\phiq{\phinoDataize{n_1}{e}{s}{\delta_1}{s}}}}{%
        \text{\phiq{\phinoNormalize{b.^}{n_1}}}%
      }%
      &
      \infer{\text{\phiq{\phinoDataize{n_2}{e}{s}{\delta_2}{s}}}}{%
        \text{\phiq{\phinoNormalize{b.|x|}{n_2}}}%
      }%
    }%
  }}%
}
\end{align*}

\Cref{fig:celsius} demonstrates the sequence of transformations
  that constitutes the dataization of \cref{eq:celsius}.

\begin{figure*}
\begin{mdframed}
\centering
\iexec[maybe]{phino --pin-file=phino-version.txt dataize --output=latex --sweet --nonumber --compress --meet-prefix=dataization --sequence --quiet --hide=Q.bytes --hide=Q.number --locator=Q.@ --focus=Q.@ --meet-length=3
--margin=150 --meet-popularity=1 --symbolic=./examples/atoms.yaml ./examples/celsius.phi | sed '1d;$d' | perl _eolang/paper-phi.pl 'equation*' /dev/stdin > \phinotex/celsius-dataization.tex}
\input{\phinotex/celsius-dataization.tex}%
\end{mdframed}
\capt{Dataization of \cref{eq:celsius}, generated by phino \iexec[maybe]{phino --version}\unskip.}
\label{fig:celsius}
\end{figure*}

% SPDX-FileCopyrightText: Copyright (c) 2016-2026 Objectionary.com
% SPDX-License-Identifier: MIT

\newpage
\section{Proofs}
\label{app:proofs}

This appendix collects the proofs of the theorems stated in \cref{sec:operators}.

% SPDX-FileCopyrightText: Copyright (c) 2016-2026 Objectionary.com
% SPDX-License-Identifier: MIT

\begin{proof}[Proof of \cref{thm:contextualization-total}]
Fix an absolute \(k\) and argue by structural induction on the expression \(e\),
  establishing at once that a derivation of \(\phinoContextualize{e}{k}{e_1}\)
  exists for some \(e_1\) and that it is finite.
By the grammar of \cref{fig:ebnf} an expression is a locator ($Q$ or
  \(\phiTerminal{\upxi}\)), a terminator $T$, a formation, a dispatch, or an
  application, and its outermost constructor tells these cases apart.
The rules of \cref{fig:contextualization} match the cases one-to-one, so exactly
  one rule applies to every expression.
For $Q$, \(\phiTerminal{\upxi}\), $T$ and a formation $[[ B ]]$
  the matching rule---\nameref{r:cg}, \nameref{r:cxi}, \nameref{r:ct} or
  \nameref{r:cf}---is an axiom that yields a result in a single step, so the base
  cases hold; \nameref{r:cf} rewrites the formation to itself and never descends
  into \(B\), halting the recursion at every formation boundary.
A dispatch $e_1.\tau$ selects \nameref{r:cd}, whose single premise
  contextualizes the subject \(e_1\); an application $e_1( \tau -> e_2 )$ selects \nameref{r:ca}, whose two premises
  contextualize \(e_1\) and \(e_2\).
Every premise is a strict subexpression of its conclusion, so by the induction
  hypothesis each has a finite derivation and a result, and the chosen rule
  assembles them into a finite derivation of the whole; the argument \(k\) is
  copied wholesale at \nameref{r:cxi} and never traversed.
The recursion therefore follows the finite syntax tree and every rule is defined
  on its case, so contextualization is total on \(\mathcal{E} \times \mathcal{K}\)
  and each derivation terminates.
\end{proof}

% SPDX-FileCopyrightText: Copyright (c) 2016-2026 Objectionary.com
% SPDX-License-Identifier: MIT

\begin{proof}[Proof sketch of \cref{thm:normalization-total}]
Consider any reduction sequence from a given expression under the rules of
  \cref{fig:reduction}; we show it is finite.

Away from \nameref{r:dot}, every rule strictly shrinks the term.
The \emph{collapsing} rules---\nameref{r:dc}, \nameref{r:dd}, \nameref{r:dl},
  \nameref{r:miss}, \nameref{r:null}, \nameref{r:over}, and
  \nameref{r:stop}---rewrite their redex to the terminator $T$, and
  \nameref{r:copy} and \nameref{r:stay} each discharge one application.
Hence only finitely many non-\nameref{r:dot} steps occur before either the term
  is exhausted or a \nameref{r:dot} redex is reached.

It remains to bound the \nameref{r:dot} steps.
\nameref{r:dot} replaces a dispatch $[[ B_1, \tau -> n, B_2 ]].\tau$ by the body
  \(n\) contextualized against $[[ B_1, B_2 ]]$ and decorated with the whole
  formation under \(\phiTerminal{\uprho}\).
An infinite sequence would need infinitely many such steps, hence a
  \nameref{r:dot} step that reproduces a dispatch of an attribute on a formation
  that still binds it---a self-reproducing redex.
We argue informally that none arises, though we do not yet exhibit the
  well-founded measure that would settle it.
The body \(n\) is contextualized against the formation with the dispatched
  attribute removed, $[[ B_1, B_2 ]]$ (\cref{sec:contextualization}), so every
  \(\phiTerminal{\upxi}\) in \(n\) that denotes the enclosing formation becomes a
  formation binding neither \(\tau\) nor $^$; a self-reference
  that reaches for \(\tau\) again, whether directly as \(\phiTerminal{\upxi}.\tau\)
  or through the parent as \(\phiTerminal{\upxi}.\phiTerminal{\uprho}.\tau\),
  dispatches a missing attribute and collapses to $T$ by \nameref{r:stop},
  \nameref{r:null}, \nameref{r:miss}, or \nameref{r:dd}.
The copy of the formation stored under \(\phiTerminal{\uprho}\) is reachable only by
  dispatching \(\phiTerminal{\uprho}\), again through such a chain, which dies the
  same way.

The calculus expresses recursion in only two further ways, and neither is a
  reduction.
Firing the atom that guards a recursive object---the base-case test that decides
  whether to recur---belongs to evaluation (\cref{sec:evaluation}): a formation
  carrying an unfired $L$-asset matches no rule of \cref{fig:reduction}
  (\nameref{r:stop} excludes it), so the recursive branch never unfolds under
  normalization.
Following the $@$-decoration to an inherited attribute belongs
  to morphing (\cref{sec:morphing}): no reduction descends into
  \(\phiTerminal{\phi}\), so dispatching an absent attribute of a decorated
  formation is irreducible rather than chasing a possibly cyclic chain.

Every dispatch a term exposes is thus drawn from its finite structure and consumed
  once, which is why \nameref{r:dot} fires finitely often and the sequence is
  finite.
Every expression therefore has a normal form, unique by
  \cref{thm:normalization-confluent}.
This last step is a sketch rather than a proof: it rests on the informal argument
  above instead of an explicit well-founded measure, and it is not yet mechanized.
Supplying such a measure, and extending the Lean~4 development that certifies
  \cref{thm:normalization-confluent} to strong normalization, remains future work.
\end{proof}

% SPDX-FileCopyrightText: Copyright (c) 2016-2026 Objectionary.com
% SPDX-License-Identifier: MIT

\begin{proof}[Proof of \cref{thm:normalization-confluent}]
We have proved this property mechanically in Lean~4,
  with the proof available in a public
  GitHub repository\footnote{\url{https://github.com/objectionary/proof}}.
\end{proof}

% SPDX-FileCopyrightText: Copyright (c) 2016-2026 Objectionary.com
% SPDX-License-Identifier: MIT

\begin{proof}[Proof of \cref{thm:morphing-confluent}]
Unlike normalization, whose rules may rewrite a redex at any position, morphing
  acts only at the root of its argument \(n\): every rule of \cref{fig:morphing}
  matches the whole of \(n\) against \(e\) and recurses through its premises on
  determined subterms.
It therefore suffices to show the rules are deterministic---at most one applies to
  a pair \((n, e)\), and the one that applies fixes the result---for then the
  relation they define is single-valued, which is the claim.
Determinism neither needs nor implies termination, so this holds even though
  morphing is partial (\cref{sec:morphing}).

\emph{At most one rule matches.}
By the grammar of \cref{fig:ebnf}, \(n\) is a locator ($Q$ or \(\phiTerminal{\upxi}\)),
  the terminator $T$, a formation, a dispatch, or an application, and its outermost
  constructor already separates most rules.
For $T$ only \nameref{r:dead} matches and for \(\phiTerminal{\upxi}\) only
  \nameref{r:xi}.
For $Q$ the universe decides: \nameref{r:universe} requires $e \neq Q$ and
  \nameref{r:mg} requires $e = Q$, so exactly one fits a given \(e\).
A formation $[[ B ]]$ with no head operation matches only \nameref{r:mf}.
A dispatch $m.\tau$ is shared by \nameref{r:ml}, \nameref{r:mphi}, and
  \nameref{r:md}, which their conditions render disjoint: \nameref{r:md} demands a
  non-formation subject (\(\phinoNotFormation{m}\)), whereas \nameref{r:ml} and
  \nameref{r:mphi} take a formation one, and among those \nameref{r:ml} requires an
  $L$-asset while \nameref{r:mphi} forbids one ($L \notin B$).
The formation-headed dispatches left over---a subject with neither an $L$-asset nor
  an $@$-attribute, or one already binding \(\tau\)---match no rule, but they are
  never normal forms: normalization reduces them by \nameref{r:stop} or
  \nameref{r:dot} (\cref{sec:normalization}), and morphing is only ever applied to a
  normal form, so it never meets them.
An application splits two ways on its argument, an exhaustive and disjoint
  partition: absolute or not (\(a \in \mathcal{K}\) for \nameref{r:ma},
  \(a \notin \mathcal{K}\) for \nameref{r:mad}), and every argument falls in
  one class.

\emph{The matching rule fixes the result.}
Its meta-variables are pinned by the term: a formation holds at most one $L$-asset,
  so \nameref{r:ml} splits $[[ B_1, L> F, B_2 ]]$ uniquely, and the subject and
  attribute of a dispatch, and the subject and argument of an application, are read
  off the syntax.
Each premise is then a single-valued judgment.
A normalization premise yields the unique normal form guaranteed by
  \cref{thm:normalization-total,thm:normalization-confluent}.
The evaluation premise of \nameref{r:ml} calls the function attached to the
  $L$-asset, which by \cref{sec:evaluation} maps a formation, the universe, and a
  state deterministically to an expression and a state, and then normalizes that
  expression to a unique normal form.
Every morphing premise is a shorter derivation, single-valued by the induction
  hypothesis.
So the rule maps its configuration, state included, to one result.

By induction on the derivation, then, \(\phinoMorph{n}{e}{s_1}{n_2}{s_2}\)
  determines \(n_2\) and \(s_2\) from \(n\), \(e\), and \(s_1\): a normal form morphs
  to at most one formation or $T$.
\end{proof}

% SPDX-FileCopyrightText: Copyright (c) 2016-2026 Objectionary.com
% SPDX-License-Identifier: MIT

\begin{proof}[Proof of \cref{thm:dataization-progress}]
By the grammar of \cref{fig:ebnf}, every expression, and hence every normal form,
  is a formation, an application, a dispatch, a locator, or the terminator $T$;
  neither the universe \(e\) nor the state \(s_1\) restricts which rule fires.
We show that every case other than $T$ matches a rule, while $T$ matches none.

If $n = T$, no rule of \cref{fig:dataization} applies, since every rule demands
  a formation or, for \nameref{r:norm}, an expression distinct from $T$.
The derivation halts with no datum, which is the error we intend: $T$ carries
  no data to read.

If \(n\) is neither a formation nor $T$,
  then \(\phinoNotFormation{n}\) and $n \neq T$,
  which is exactly the condition of \nameref{r:norm};
  the rule morphs \(n\) and dataizes the result,
  and since morphing delivers a formation or $T$ (\cref{sec:morphing}),
  that result falls into one of the other cases.

If $n = [[ B ]]$ is a formation, we inspect its assets and attributes in order:
  \begin{inparaenum}[a)]
  \item if \(B\) carries a $D$-asset, \nameref{r:delta} fires;
  \item otherwise, if \(B\) carries an $L$-asset, \nameref{r:fire} fires;
  \item otherwise, if \(B\) carries an $@$-attribute, \nameref{r:box} fires,
    its condition $[D, L] \cap B = \emptyset$ granted by the two exclusions above;
  \item otherwise \(B\) holds none of $D$, $L$, and $@$,
    which is the condition of \nameref{r:none}; that rule fires and reduces
    the goal to the dataization of $T$, which then halts as above --- a
    formation with nothing to dataize is itself an error.
  \end{inparaenum}
So every formation matches a rule, and $T$ alone is left without one.

Termination is by induction on the derivation.
\nameref{r:delta} is an axiom and discharges with no premise.
Each remaining rule reduces the dataization of \(n\) to the dataization of a normal
  form produced by an operator that itself terminates---contextualization
  (\cref{sec:contextualization}) and normalization (\cref{sec:normalization})
  for \nameref{r:box}, evaluation and normalization for \nameref{r:fire},
  and morphing (\cref{sec:morphing}) for \nameref{r:norm}, while \nameref{r:none}
  reduces directly to $T$---so every premise is again a dataization judgment,
  and \nameref{r:norm} never recurs into itself.
The derivation is hence finite whenever the recursion through \nameref{r:box}
  and \nameref{r:fire} is well-founded,
  which holds for every expression that denotes a terminating computation;
  such a derivation ends either at a datum or at the terminator $T$.
\end{proof}

% SPDX-FileCopyrightText: Copyright (c) 2016-2026 Objectionary.com
% SPDX-License-Identifier: MIT

\begin{proof}[Proof of \cref{thm:dataization-confluent}]
Dataization is specified by the syntax-directed inference system of
  \cref{fig:dataization}, so confluence amounts to the relation $\Downarrow$
  being single-valued: whenever \(\phinoDataize{n}{e}{s_1}{\delta}{s_2}\)
  and \(\phinoDataize{n}{e}{s_1}{\delta'}{s_2'}\), we have \(\delta = \delta'\)
  and \(s_2 = s_2'\).
We argue by induction on the first derivation, after observing that at most
  one rule applies to any configuration \(\langle n, e, s_1 \rangle\),
  so both derivations are headed by the same rule.

The conclusions of the five rules partition every configuration, and $T$ falls
  under none of them.
If $n = T$, no rule matches, so dataization is undefined and there is nothing
  to disprove.
If \(n\) is neither a formation nor $T$, only \nameref{r:norm} matches.
If $n = [[ B ]]$ is a formation, the asset and attribute it carries select
  a single rule:
  \begin{inparaenum}[a)]
  \item a $D$-asset selects \nameref{r:delta} and an $L$-asset selects
    \nameref{r:fire}, the two being mutually exclusive in a normal form;
  \item with neither asset present, an $@$-attribute selects \nameref{r:box}
    and its absence selects \nameref{r:none},
  \end{inparaenum}
  the latter pair separated by the explicit conditions
  $[D, L] \cap B = \emptyset$ and $[ D, L, @ ] \cap B = \emptyset$.
Hence no configuration matches two rules.

Because the same rule heads both derivations, it suffices to check that each
  rule maps its premises to a single result.
\nameref{r:delta} is an axiom whose conclusion reads the result directly off
  \(n\)---the attached data---leaving the state untouched, so it admits no other
  outcome.
\nameref{r:none} reduces to the dataization of $T$, which is undefined, so it
  yields no result and hence no second one.
\nameref{r:box} chains three premises: contextualization is a total function
  (\cref{sec:contextualization}) and yields a unique \(n\); normalization is
  confluent (\cref{sec:normalization}) and yields a unique normal form \(n_1\);
  and the dataization of \(n_1\) is unique by the induction hypothesis.
\nameref{r:norm} chains morphing, which is confluent (\cref{sec:morphing}) and
  yields a unique formation or $T$, with a dataization premise that is unique
  by the induction hypothesis.
\nameref{r:fire} evaluates the function \(F\) attached to the $L$-asset.
By \cref{sec:evaluation} a function is a total mapping that deterministically
  maps formation, universe, and state to an expression and a new state,
  so its result depends on nothing else.
The universe \(e\) is fixed throughout the derivation and the state propagates
  only through the explicit threading of the rules, so both derivations present
  the call with identical arguments and it returns the same
  \(\langle n, s_2 \rangle\); normalization then yields a unique \(n_1\) and the
  dataization premise is unique by the induction hypothesis.
In every case the preceding premises are determined before the recursive one,
  so the sub-derivations dataize the same normal form in the same state and the
  induction hypothesis applies.
The two derivations therefore agree on the result, and dataization relates each
  normal form to at most one datum and state.
\end{proof}

\end{document}